\documentstyle[11pt,emulateapj]{article}
\tighten

\begin{document}

\def\beginrefer{\section*{References}%
\begin{quotation}\mbox{}\par}
\def\refer#1\par{{\setlength{\parindent}{-\leftmargin}\indent#1\par}}
\def\endrefer{\end{quotation}}

\title{ Chandra Observations of
the Gravitationally Lensed System 2016+112}

\author{G. Chartas\altaffilmark{1}, M. Bautz\altaffilmark{2}, 
G. Garmire\altaffilmark{1}, C. Jones\altaffilmark{3}, and D. P. Schneider\altaffilmark{1}} 

\altaffiltext{1}{Astronomy and Astrophysics Department, The Pennsylvania
State University, University Park, PA 16802.}
\altaffiltext{2}{MIT Center for Space Research, 70 Vassar Street, Cambridge, MA, 02139.}
\altaffiltext{3}{Harvard-Smithsonian Center For Astrophysics, Cambridge, MA 02138}
\begin{abstract}
An observation of the gravitationally lensed
system 2016+112 with the Chandra X-ray Observatory has resolved
a mystery regarding the proposed presence of a dark matter object
in the lens plane of this system.
The {\it Chandra} ACIS observation has clearly detected the lensed
images of 2016+112 with positions in good agreement
with those reported in the optical and also detects
13 additional X-ray sources within a radius of 3.5 arcmin.
Previous X-ray observations in the direction of 2016+112
with the ROSAT HRI and ASCA SIS have 
interpreted the X-ray data as arising from extended
emission from a dark cluster.
However, the present {\it Chandra} observation
can account for all the X-ray emission
as originating from the lensed images and additional point X-ray
sources in the field. Thus cluster parameters based 
on previous X-ray observations are unreliable. 
We place a 3$\sigma$ upper limit on the 2-10keV
flux and luminosity of the cluster of
1.6 $\times$ 10$^{-14}$ erg s$^{-1}$ cm$^{-2}$ and 
1.7 $\times$ 10$^{44}$ erg s$^{-1}$, respectively.
We estimate an upper limit on the  mass-to-light ratio within a radius
of 800 $h_{50}$$^{-1}$ kpc of $M$/L$_{V}$ $ < $ 190 $h_{50}$ $(M/L_{V})$$_{\odot}$.
None of the additional point X-ray
sources are associated with the galaxy cluster members
recently detected in deep optical and IR observations (Soucail et al. 2000).

The lensed object is quite unusual, with reported narrow emission lines in the
optical that suggest it may be a type-2 quasar (Yamada et. al. 1999).
Our modeling of the X-ray spectrum of the lensed object
implies that the column density of an intrinsic absorber must lie
between 3 and 85 $\times$ 10$^{22}$ cm$^{-2}$ (3$\sigma$ confidence level).
The 2-10~keV luminosity of the lensed object, corrected 
for the lens magnification effect and
using the above range of intrinsic absorption, is
3 $\times$ 10$^{43}$  - 1.4 $\times$ 10$^{44}$ erg s$^{-1}$.

\end{abstract}


\section{INTRODUCTION}

Wide-separation bona fide gravitational lens (GL) systems  
with spectroscopically confirmed lens redshifts
have been found to contain, in many cases, galaxy-cluster or
galaxy-group lenses. The presence of such massive deflectors contributes
significantly to the large deflection angles observed in these systems.
Solid cases of wide image separation systems containing galaxy cluster
lenses are RXJ0911+0551 (Burud et al. 1998) and Q0957+561 
(Walsh, Carswell \& Weyman 1979; Garrett et al. 1992; Angonin et al. 1994). 
2016+112 also falls in the category 
of confirmed wide-separation GL systems
and was first identified by Lawrence et al. (1984)
to be comprised of at least three lensed images of an AGN at a redshift
of 3.273. The lens contains a giant elliptical galaxy 
at $z$ = 1.01 (Schneider et al. 1986; 
Langston, Fisher, \& Aspin 1991; Lawrence, Neugebauer, \& Matthews 1993).
The initial deep optical observations of 
2016+112 (Schneider et al. 1985; 1986; Lawrence, Neugebauer, \& Matthews 1993)
did not reveal a galaxy cluster near the lens position.

Hattori et al. (1997) reported the detection of 
resolved X-ray emission at $z$ $\sim$ 1
centered on the giant elliptical.
This result was based in part on the detection of an X-ray line
at 3.35~keV in the ASCA SIS spectra of 2016+112;
the feature was interpreted as Fe-K line emission originating from a 
cluster at $z$ $\sim 1$.
One of the significant conclusions of the Hattori et al. (1997) detection
was the unexpectedly large M/L ratio $\sim$ 3000
implied by the X-ray luminosity 
(L$_{X}$(2-10~keV) = 8.5 $\times$ 10$^{44}$  erg s$^{-1}$)
and the blue luminosity of the giant elliptical 
lens galaxy (L$_{B}$ = 1.1 $\times$ 10$^{11}$ $L_{\odot}$).
Their spectral fits to the ASCA SIS spectra of 2016+112
also imply a high iron abundance, Z = 1.7 $\times$ $Z_{\odot}$.
One of the puzzling aspects of the Hattori et al. (1997)
result was the non-detection of optical and IR emission
from the stars in the cluster galaxies even though the 
X-ray data suggested the presence of metal rich galaxies (Mushotsky et al. 1997). 
The spatial resolution of $\sim$ 3~arcmin and positional
accuracy of $\sim$ 1 arcmin of ASCA were inadequate to
establish whether the detected X-ray emission was extended.

Detection of extended X-ray emission near the giant elliptical galaxy
was needed to confirm the presence of a galaxy cluster, so
2016+112 was observed with the ROSAT HRI,
which has a spatial resolution of $\sim$ 5~arcsec.
Hattori et al. (1997) detected
 a ``diffuse'' source with 76  $\pm$ 24 counts (0.2 - 2~keV),
(1.4 $\pm$ 0.4) $\times$ 10$^{-3}$ counts s$^{-1}$,
within 1 arcmin radius. The Hattori et al. (1997) analysis suggested that the
object was extended.
Benitez et al. (1999) re-analyzed the ROSAT HRI data
and found a faint, elliptical X-ray source spatially
coincident with a red galaxy over-density. In particular,
their deep imaging in $V$ and $I$ bands using Keck identified
a red sequence of galaxies in color-magnitude diagrams of $V-I$ vs. $I-K$,
suggesting that these are members of a galaxy cluster at $z$ $\sim$ 1.
Additional spectroscopic observations have 
confirmed the presence of several red galaxy members at a redshift $z$ $\sim$ 1
(Kneib et al. 1998; Soucail et al. 2000; Clowe et al. 2000).
This faint X-ray source is considered in their analysis as the
counterpart of the optical galaxy cluster. 
Two additional X-ray sources are detected in the vicinity of 2016+112
that were smoothed together in the analysis by Hattori et al. (1997). 

The Benitez et al. (1999) analysis indicated that the cluster lens
is an order of magnitude more luminous in the optical than previously
reported and the X-ray luminosity is lower than
previously estimated. A revised mass-to-light ratio 
of M/L$_{V}$ = 372 $\pm$ 94 $h_{100}$(M/L$_{V}$)$_{\odot}$ 
, within a radius of 400 $h_{100}^{-1}$ kpc, for the cluster
lens of 2016+112 was reported by Benitez.
The hypothesis of the presence of
a dark matter lens in 2016+112 is not supported with
the Benitez et al. (1999) analysis and the recent 
Chandra observation presented in this letter.

With the launch of the {\it Chandra} X-ray observatory on
23 July 1999, it has been possible for the first time
to spatially resolve the components of most GL systems in the X-ray band.
The GL system 2016+112  was
observed with {\it Chandra} as part of a snapshot 
survey of GL quasars aimed at finding
suitable candidates for time-delay measurements.

In this letter we describe our observation of 2016+112
using the {\it Chandra} Advanced CCD Imaging Spectrometer 
(ACIS; Garmire et al. 2000, in preparation), 
results from our data analysis and the comparison to previous X-ray 
results. 
We use $H_{0}$ = 50 km s$^{-1}$ Mpc$^{-1}$,
q$_{0}$ = 0.5, and $\Lambda$ = 0, unless mentioned otherwise.

\section{CHANDRA OBSERVATIONS AND DATA ANALYSIS}

The GL system 2016+112 was observed with {\it Chandra} on April 12, 2000
for $\sim$ 7.7 ks. The X-ray image of the {\it Chandra} observation is 
shown in figure 1a. A comparison of this image with that 
taken with the ROSAT HRI has resolved 
the mystery surrounding the reported detection
of extended X-ray emission that was attributed to a dark cluster of galaxies
at a redshift of z $\sim$ 1. With the {\it Chandra} 
observation, we clearly 
identify several point-sources in this region and 
detect the three lensed images, A, B, and C (see figure 1b). 
The image presented in figure 1b was obtained by binning the 
X-ray event positions to 0.1$''$ and smoothing the binned image 
with a Gaussian of width $\sigma$ = 0.2$''$.

The X-ray image of 2016+112 obtained by the ROSAT HRI and presented 
in Hattori et al. (1997) is shown in figure 2a. 
In figure 2b, we show the {\it Chandra} image of 2016+112 rebinned by a factor of 12
and smoothed with a Gaussian of width $\sigma$ = 15$''$, 
to reproduce the Hattori el al. 1997 analysis of the ROSAT HRI image.
It is obvious that the extended component in the 
ROSAT HRI observation was merely an artifact produced by
the inadequate spatial resolution of the ROSAT HRI and image smoothing.
The unbinned {\it Chandra} image shows several point sources in the 
vicinity of 2016+112. {\it Chandra} has resolved the image components with 
locations in good agreement with recently reported optical and 
radio positions for this object. Table 1 lists the relative 
positions of the lensed images B and C
with respect to A as observed with {\it Chandra} and HST.
The properties of the detected sources in the
vicinity of 2016+112 are presented in Table 2.
 
The reported ROSAT HRI count-rate for 2016+112 of 
1.4 $\pm$ 0.4 $\times$ 10$^{-3}$ cnts s$^{-1}$ (Hattori et al. 1997)
was extracted within a 1 arcmin circle centered on 2016+112 
and would have included the three additional point sources 
resolved with {\it Chandra} (see figure 2). We converted the observed ROSAT HRI and 
ASCA SIS count-rates of 2016+112 to ACIS-S S3 CCD count-rates 
of 1.9 $\pm$ 0.6 $\times$ 10$^{-2}$ cnts s$^{-1}$ and 
3.4 $\pm$ 0.3 $\times$ 10$^{-2}$cnts s$^{-1}$, respectively,  
assuming an absorbed power law model with a Galactic
column density of 15.6 $\times$ 10$^{20}$ cm$^{-2}$ and 
a photon spectral index of 2. The total count-rate observed with the 
ACIS-S S3 CCD from the lensed images 
of 2016+112 and three additional point sources within a 
circle of 1 arcmin radius centered on 2016+112 is 
1.4 $\pm$ 0.2 $\times$ 10$^{-2}$cnts s$^{-1}$, consistent
with the ROSAT HRI count-rate. The total count-rate 
detected with ACIS from the lensed images
of 2016+112 and additional point sources within a
circle of 3.5~arcmin radius centered on 2016+112 is 
3.3 $\pm$ 0.2 $\times$ 10$^{-2}$cnts s$^{-1}$ consistent 
with the ASCA count-rate.

As an additional comparison between the ASCA and {\it Chandra} results,
we extracted a composite spectrum from only the point-like
sources located within a circle of 3.5~arcmin radius of 2016+112.
Following the standard Chandra X-ray Center (CXC) analysis procedures,
we applied the appropriate gain map correction, eliminated times with bad aspect
and created response and anciliary files using the CXC software tools
{\it mkrmf} and {\it mkarf}. 
During the {\it Chandra} observation of 2016+112 the background detected on the 
ACIS CCD S3 was fairly constant with a rate of 
approximately 4.4 $\times$ 10$^{-7}$ counts s$^{-1}$ pixel$^{-1}$
within the energy band 0.4 to 8~keV and for events with standard
ASCA grades 0,2,3,4,6. Our simple spectral fit to the {\it Chandra} 
composite spectrum of all point sources within 3.5 arcmin 
centered on 2016+112 yields a 2-10~keV X-ray luminosity of 
L$_{X}$ = 9.0$^{+0.3}_{-0.3}$ $\times$ $10^{44}$ $h^{-2}_{50}$ erg s$^{-1}$. 
The 2-10~keV X-ray luminosity reported for the ASCA observation of 2016+112
is L$_{X}$ = 8.4$^{+2.4}_{-1.7}$ $\times$ $10^{44}$ $h^{-2}_{50}$ erg s$^{-1}$.
We conclude that the X-ray flux attributed to
a galaxy cluster in the lens plane of   
the 2016+112 GL system is accounted for by point-like X-ray sources
resolved by {\it Chandra}.

We compared the locations of the X-ray sources detected with {\it Chandra} 
in the field of 2016+112 (Table 2) to those detected in the optical 
with the Nordic Optical Telescope (Table 1 in Soucail et al. 2000). We find no
matches within 5$'$ of 2016+112, indicating that the detected 
X-ray sources are not associated with 
the galaxies detected by Soucail et al. (2000)
that are bound to the $z$ = 1.01 cluster.
We also searched the USNO catalog and found no optical counterparts
to the X-ray sources.

\section{CLUSTER MASS AND MASS-TO-LIGHT ESTIMATES}

The lenses of several candidate gravitationally lensed systems
with image separations larger than $\sim$ 3 arcsec
have eluded detection in the optical and radio bands. 
These non-detections have lead to the hypothesis
of the existence of large dark matter lenses
with mass-to-light ratios $M/L  > 1000(M/L)_{\odot}$
(Schneider, Ehlers, and Falco 1992). 
The GL system 2016+112 was originally proposed as 
containing a dark matter lens based on the X-ray mass estimate of 
3.6$_{-1.3}^{+1.8}$ $\times$ 10$^{14}$ $h_{50}^{-1}$ M$_{\odot}$ within 500 $h_{50}^{-1}$ kpc
(Hattori et. al. 1997) and the blue luminosity of the lensing galaxy D.

As described in the introduction, recent deep optical and IR 
observations have detected several
galaxy cluster members, significantly revising the luminosity
of the cluster. Specifically, the reported value
for the luminosity of the galaxy cluster members 
is 3.2  $\times$ 10$^{11}$ $h_{100}^{-2}$ L$_{{V}{\odot}}$ 
within a radius of 400 $h_{100}^{-1}$ kpc (Benitez et. al. 1999).
The significant galactic extinction in the
direction of 2016+112 has made it extremely difficult to obtain
reliable estimates for the luminosities of the galaxy cluster members,
even in the $I$ band. 
By applying the cluster virial theorem to 6 galaxies of the cluster
Soucail et al. (2000) find a virial mass of 
$M$ = 2.8$^{+4.0}_{-1.0}$ $\times$ 10$^{14}$ $h_{50}^{-1}$ M$_{{\odot}}$
within a radius of 350 $h_{50}^{-1}$ kpc.

Our {\it Chandra} observations indicate that the cluster mass estimates 
based on previous X-ray measurements must be revised.
In particular,  {\it Chandra} has clearly resolved the three X-ray
lensed images of 2016+112, and 13 additional point-like
sources within 3.5 arcmin centered on the lensed AGN.
Our estimated X-ray 2-10~keV fluxes within circles 
of radii 1 and 3.5~arcmin centered on 2016+112 and 
originating from point-like sources alone
are 0.7 $\pm$ 0.2 $\times$ 10$^{-13}$ erg s$^{-1}$ cm$^{-2}$ and 
1.3 $\pm$ 0.1 $\times$ 10$^{-13}$ erg s$^{-1}$ cm$^{-2}$, 
in good agreement with the reported values from
similar sized regions of ROSAT HRI and ASCA.
The ASCA spectrum of this source was 
therefore dominated by emission from point-like objects 
unrelated to the cluster. Any cluster parameters
derived from the ASCA spectral analysis cannot be considered reliable.
Also previous estimates of the cluster core radius and $\beta$ parameter
based on the ROSAT HRI observations are incorrect.
Obviously, the previously reported total mass of the 
cluster lens of 2016+112
derived from cluster parameters $kT$, r$_{c}$
and $\beta$ must be considered unreliable.

To estimate the upper limit on the X-ray luminosity
of an extended component we extracted events 
within a 1.56 arcmin radius ( $\sim$ 0.8~Mpc)
centered on 2016+112 and excluded all point sources within 
this region listed in Table 2. We used an iterative procedure to place 
3$\sigma$ upper bounds on the intracluster temperature and 
2-10~keV luminosity. We began with a initial
guess of 3~keV for the cluster temperature and estimated
the X-ray luminosity assuming a Raymond-Smith
thermal spectrum for the ICM gas, a redshift of $z$ = 1.01,
an abundance of 0.3 cosmic, and a column density (Galactic)
of N$_{H}$ = 0.15 $\times$ 10$^{22}$ cm$^{-2}$.
From the $L_{X} - T$ relation for clusters of galaxies (Markevitch, 1998), we 
select a new trial cluster temperature and repeat the 
calculation of the cluster luminosity.
We repeat this iterative process untill the solution
converges. We place a 3$\sigma$ upper
limit of 3.7~keV on the cluster temperature and
3$\sigma$ upper limits of 1.7 $\times$ 10$^{44}$ erg s$^{-1}$
and 1.6 $\times$ 10$^{-14}$ erg s$^{-1}$ cm$^{-2}$ on the 2-10~keV luminosity and flux of the cluster.

Using the empirically derived relation between total cluster mass
and ICM temperature, M$_{500}$ = 2.00$\times$ 10$^{15}$ $h^{-1}_{50}$ M$_{\odot}$ (T$_{X}$/10~keV)$^{3/2}$,
where M$_{500}$  is the mass within the radius r$_{500}$, in which the mean overdensity is 500,
and r$_{500}$ = 2.37 $h^{-1}_{50}$ Mpc (T$_{X}$/10~keV)$^{1/2}$,  
(Mohr, Mathiesen, \& Evrard, 1999)  
we find that our temperature limit of 3.7~keV
corresponds to a mass limit of,
M$_{500}$ $ < $ 4.5 $\times$ 10$^{14}$ $h^{-1}_{50}$ M$_{\odot}$ 
with r$_{500}$ = 1.44 $h_{50}^{-1}$ Mpc. 
We use a luminosity from all galaxy types of
L$_{V}$ = 1.28  $\times$ 10$^{12}$ $h_{50}^{-2}$ L$_{{V}{\odot}}$ 
within a radius of 800 $h_{50}^{-1}$ kpc, as derived by Benitez et. al. (1999),
and our estimated upper-bound on the cluster mass to derive
a mass-to-light ratio limit of M/L $ < $ 190 $h_{50}$ (M$_{\odot}$/L$_{{V}{\odot}}$)
within 800 $h_{50}^{-1}$ kpc.

Our 7.7~ks {\it Chandra} observation does not
yield any useful upper limit on emission
from a possible redshifted Fe-K$\alpha$ line 
originating from the cluster of galaxies. 
A much deeper observation of 2016+112 with the ACIS-I CCD
that has lower background than the ACIS-S CCD may result in the detection and
characterization of the properties of the distant z=1.01 cluster of galaxies.  
In addition to the cosmological implications of detecting
clusters of galaxies at high redshifts, an estimate of
the center of mass and mass distribution will
improve gravitational lens models of this system.
These improved models may yield an estimate for the Hubble constant
and also provide the absolute distance to the lensing cluster.

\section{X-RAY PROPERTIES OF LENSED AGN}

The optical spectra of the lensed images of 2016+112 are quite unusual;
strong and narrow emission lines of Ly$\alpha$, N${\bf V}$, C${\bf IV}$, 
He${\bf II}$ and C${\bf III]}$ at a redshift of 3.273 are  
detected in all images (Lawrence et al. 1984; Yamada et al., 1999). 
It has been recently proposed that the lensed object 
is a luminous type-2 quasar (Yamada et al. 1999).
Here we define as a type-2 quasar an AGN that
contains no broad optical emission-lines and is sufficiently luminous 
(i.e., L$_{X}$ (2-10~keV) $\stackrel{>}{\sim}$ 2 $\times$ 10$^{44}$ erg s$^{-1}$ )
to qualify as a quasar.
We investigated the nature of the lensed object by
estimating its unabsorbed X-ray luminosity.
Unfortunately, the lensed AGN is quite faint in X-rays with
6, 5, and 12 detected counts in images A, B and C respectively.
The similarity of the optical and radio spectra of images A and B 
suggests that they are produced from the gravitationally lensing
of a background AGN by the giant elliptical galaxy D and the host 
cluster of galaxies. 
Yamada et al. (1999) conclude that the optical spectra 
of images A, B and C are consistent in showing a type-2 spectrum.
Yamada et al. (1999) attribute the observed differences 
in the optical emission line ratios
in images B and C to the different degree of photoionization of the lensed regions
that produce the emission of these images.
In particular, recent lens modeling of 2016+112 (Benitez et al. 1999) can 
reproduce the observed morphology, including the extended 
nature of image C, by placing the center of the lensed source just outside  
the magnification caustic and an extended emission component 
within the caustic. 
The optical emission in image C, thus, may arise
from larger radii of the narrow line region (NLR), its radio emission 
from a possible jet, and its X-ray emission may be lensed 
synchrotron self Compton (SSC) emission from the jet.
Since we are interested in estimating the unlensed luminosity of
the central object we only consider the emission from 
images A and B.
We modeled the {\it Chandra} spectrum of 2016+112 images A and B
with an absorbed power-law  with 
the Galactic contribution to the column density held 
fixed at N$_{H}$ = 0.15 $\times$ 10$^{22}$ cm$^{-2}$,
an intrinsic absorption column at $z$  = 3.27 and a spectral slope
fixed at 2. The 90\% confidence range for the intrinsic 
absorption column is (3  - 85) $\times$ 10$^{22}$ cm$^{-2}$.
The luminosity in the rest frame 2-10~keV band 
corrected for lens magnification for the above 
range of intrinsic absorption is 
3 $\times$ 10$^{43}$  - 1.4 $\times$ 10$^{44}$ erg s$^{-1}$.
We assumed magnification factors of 8.5 and 6 for the observed X-ray images
A and B respectively, based on the lens models of Benitez et al. (1999).
Our upper bound on the X-ray luminosity of the lensed object places it
near the ``grey'' area between low and high luminosity AGN's.

\section{CONCLUSIONS}

A {\it Chandra} observation of the GL system 2016+112
has clearly resolved the image components of this
system in agreement with previous optical and radio observations.
Derived parameters for the mass and M/L of
the lensing cluster based on previous X-ray observations are
unreliable due to the contamination produced by several 
X-ray sources in the field. 
Based on the non-detection of the lensing cluster of galaxies
in this 7.7ks Chandra observation we place a 3$\sigma$ upper limit
of 1.7 $\times$ 10$^{44}$ erg s$^{-1}$ on the 2-10~keV luminosity of the cluster.
We place upper limits on the mass and mass-to-light
ratio of the galaxy cluster of $M$ $ < $ 2.5 $\times$ 10$^{14}$ $h^{-1}_{50}$ M$_{\odot}$ 
and $M$/L$_{V}$ $ < $ 190 $h_{50}$ $(M/L_{V})$$_{\odot}$
within 800 $h_{50}$$^{-1}$ kpc, respectively.
With the detection of galaxy members of the cluster in the optical band 
(Soucail et al., 2000; Benitez et al. 1999) 
and the present revision of the cluster X-ray luminosity, 
we conclude that the lensing cluster of 2016+112 
can be ruled out as a dark cluster candidate.
The nature of the lensing source is still unclear.
We estimate that the X-ray luminosity of the 
lensed object (corrected for intrinsic absorption and the magnification effect)
must be less than 1.4 $\times$ 10$^{44}$ erg s$^{-1}$ at
the 3$\sigma$ level which is within the range of a Seyfert galaxy.\\

We would like to thank M. Eracleous for helpful discussions and comments.
We acknowledge financial support by NASA grant NAS 8-38252
and AST99-00703 (DPS).

\newpage

\small

\begin{center}
\begin{tabular}{llllll}
\multicolumn{6}{c}{TABLE 1}\\
\multicolumn{6}{c}{Optical and X-ray Offsets of 2016+112 Images} \\
& & & & & \\ \hline\hline
\multicolumn{1}{c} {Telescope} &
\multicolumn{1}{c} {(RA, Dec)$_{A}$ ${}^{a}$ } &
\multicolumn{1}{c} {(RA, Dec)$_{B}$ ${}^{a}$ } &
\multicolumn{1}{c} {(RA, Dec)$_{C}$ ${}^{a}$ } &
\multicolumn{1}{c} {B/A ${}^{b}$ } &
\multicolumn{1}{c} {C/A ${}^{b}$}  \\
        &arcsec& arcsec        &  arcsec        &        &           \\ \hline
HST     &(0,0) &(-3.010,-1.506)&(-2.057,-3.255) &0.97 $\pm$ 0.02     &1.20 $\pm$ 0.04  \\
CHANDRA &(0,0) &(-2.9,-1.2)    &(-2.0,-3.2)     &0.8 $\pm$ 0.6 &2 $\pm$ 1.3 \\
\hline \hline
\end{tabular}
\end{center}

NOTES-\\
${}^{a}$ Offsets in RA and Dec with respect to image A \\
${}^{b}$ Image flux ratios.
The {\it Chandra} and {\it HST} flux ratios are calculated in the 0.4-8~keV
and H bands (centered on 1600 \AA), respectively.
The {\it HST} offsets and flux ratios are taken from the 
CfA-Arizona Space Telescope LEns Survey (CASTLES) of gravitational lenses 
website {\it http://cfa-www.harvard.edu/glensdata/}.

\normalsize

\newpage

\scriptsize

\begin{center}
\begin{tabular}{lllrr}
\multicolumn{5}{c}{TABLE 2}\\
\multicolumn{5}{c}{Sources in the Near Vicinity of 2016+112} \\
& & & &  \\ \hline\hline
\multicolumn{1}{c} {Object} &
\multicolumn{1}{c} {RA (J2000)} &
\multicolumn{1}{c} {DEC (J2000)} &
\multicolumn{1}{c} {Distance}$^{a}$ &
\multicolumn{1}{c} {Source Counts}$^{b}$  \\
    &       &        & arcsec            &    cnts             \\ \hline
CXO J201917.3+112751  &20 19  17.37  &11  27 51.8 & 39.9 &42.9 $\pm$  6.5 \\
CXO J201920.1+112627  &20 19  20.13  &11  26 27.9 & 54.2 &30.8 $\pm$  5.6\\
CXO J201917.7+112555  &20 19  17.78  &11  25 55.7 & 78.0 &23.9 $\pm$ 4.9 \\
CXO J201921.3+112831  &20 19  21.36  &11  28 31.6 & 91.4 &17.9 $\pm$ 4.2 \\
CXO J201915.5+112458  &20 19  15.58  &11  24 58.5 &140.2 &34.6 $\pm$ 5.9 \\
CXO J201916.0+112454  &20 19  16.01  &11  24 54.8 &142.2 &10.9 $\pm$ 3.3 \\
CXO J201929.0+112855  &20 19  29.06  &11  28 55.0 &190.2 &14.8 $\pm$ 3.9 \\
CXO J201921.6+112356  &20 19  21.62  &11  23 56.3 &203.8 &17.8 $\pm$ 4.2 \\
CXO J201908.7+112434  &20 19  8.782  &11  24 34.3 &210.3 &12.8 $\pm$ 3.6 \\
CXO J201930.3+112450  &20 19  30.34  &11  24 50.0 &229.8 &42.3 $\pm$ 6.6 \\
CXO J201929.1+112931  &20 19  29.11  &11  29 31.9 &212.5 &14.9 $\pm$ 3.9 \\
CXO J201931.4+112902  &20 19  31.49  &11  29 02.2 &224.3 & 9.8 $\pm$ 3.2 \\
CXO J201922.3+112227  &20 19  22.35  &11  22 27.5 &292.6 &11.4 $\pm$ 3.5 \\
\hline \hline
\end{tabular}
\end{center}

NOTES-\\
${}^{a}$ Distance from image 2016+112 C.\\
${}^{b}$ Source counts extracted with the CXC software tool {\it wavdetect}
in the 0.4-8~keV band.

\normalsize

\newpage

\normalsize

\beginrefer
\refer Angonin-Willaime, M.-C., Soucail, G., Vanderriest, C., 1994, A\&A, 291, 411 \\

\refer Benitez, N., Broadhurst, T., Rosati, P., Courbin, F., 
Gordon, S., Lidman, C., \& Magain, P., 1999, ApJ, 527, 31 \\

\refer {Burud}, I., {Courbin}, F., {Lidman}, C., {Jaunsen}, A. O.,
{Hjorth}, J., {Ostensen}, R., {Andersen}, M. I., {Clasen}, J. W.,
{Wucknitz}, O., {Meylan}, G., {Magain}, P., {Stabell}, R.,
and {Refsdal}, S., 1998, \apjl, 501, L5 \\

\refer Clowe, D., Trentham, N., and Tonry, J., 2000, 
A\&A, astro-ph/0001309 \\

\refer Garrett, M. A., Walsh, D., and Carswell, R. F., 1992, MNRAS, 254, 27 \\ 

\refer Hattori, M., Ikebe, Y., Asaoka, I., Takeshima, T., Bohringer, H., 
Mihara, T., Neumann, D. M., Schindler, S., Tsuru, T., \& Tamura, T., 1997,
Nature, 388, 146 \\

\refer Kneib, J.-P., Soucail, G., Jaunsen, A., Hattori, M., Hjorth, J., \&
Yamada, T., 1998, CFHT Inf. Bull. 38. \\

\refer Langston, G., Fischer, J. \& Aspin, C., 1991, AJ, 102, 1253 \\

\refer Lawrence, C. R., Neugebauer, G. \& Matthews, K. 1993, AJ, 105, 17 \\

\refer Lawrence, C. R., Schneider, D. P., Schmidt, M., 
Bennett, C, L., Hewitt, J. N., Burke, B. F., Turner, E. L., 
and Gunn, J. E., 1984, Science, 223, 46 \\

\refer Markevitch, M., 1998, $\apj$, 504, 27 \\

\refer Mohr, J. J., Mathiesen, B. \& Evrard, A. E., 1999, $\apj$, 517, 627 \\

\refer Mushotzky, R., 1997, Nature, 388, 126 \\

\refer {Schneider}, D. P., {Lawrence}, C.  R., {Schmidt}, M.,
{Gunn}, J. E., {Turner}, E. L., {Burke}, B. F. and 
{Dhawan}, V., 1985, \apj, 294, 66\\

\refer Schneider, D. P., Gunn, J. E., Turner, E. L., 
Lawrence, C. R., Hewitt, J. N., Schmidt, M. \& Burke, B. F. 1986, AJ, 91, 991 \\

\refer Soucail, G., Kneib, J.-P., Jansen, A. O., Hjorth, J., Hattori, M., 
and Yamada, T., 2000, submitted to A\&A, astro-ph/0006382 \\

\refer Yamada, T., S. Yamazaki, M. Hattori, G. Soucail, and J.- P. Kneib,
submitted to A\&A, astro-ph/9908089 \\

\refer Walsh, D., Carswell, R. F., \& Weymann, R. J., 1979, Nature, 279, 381.

\endrefer

\clearpage

\begin{figure*}[t]
\plotfiddle{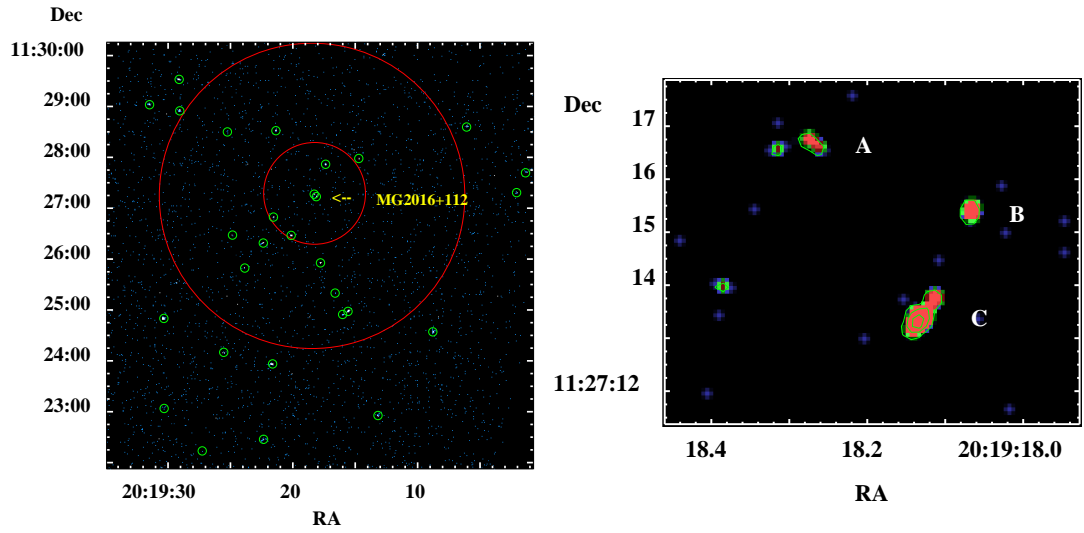}{2.in}{0}{80.}{80.}{-250}{-320}
\protect\caption 
{\small (left panel) 8$'$ $\times$ 8$'$ X-ray image (0.4 - 8~keV band)
of the 2016+112 field obtained by the {\it Chandra} X-ray Observatory.
Sources detected with the software tool {\it wavdetect} are circled.
Circles of radii 1$'$ and 3$'$ centered on 2016+112 are also overlade for
comparison to the ROSAT HRI and ASCA analysis by Hattori et al. (1997)
(right panel) Gravitationally lensed images A, B and C of the quasar 2016+112
resolved by the {\it Chandra} X-ray Observatory.
\label{fig:fig1}}
\end{figure*}

\clearpage

\begin{figure*}[t]
\plotfiddle{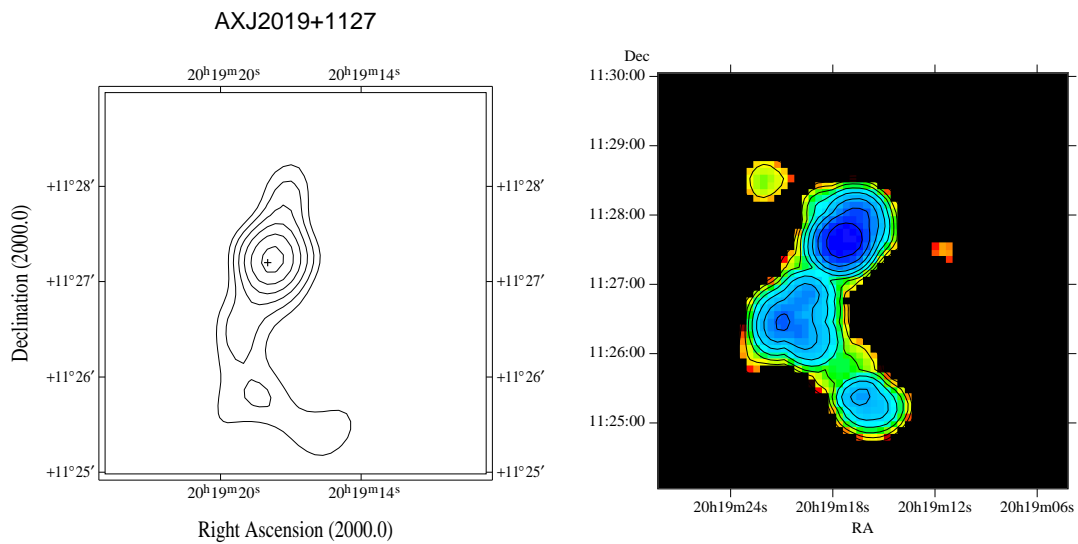}{2.in}{0}{80.}{80.}{-250}{-320}
\protect\caption 
{\small (left panel) X-ray HRI image of 2016+112 (0.2 - 2.2~keV band)
obtained by the ROSAT HRI (image from Hattori et al. 1997). The image was smoothed
in their analysis with a Gaussian filter with $\sigma$ = 15$''$.
(right panel) X-ray image of 2016+112 (0.4 - 8~keV band)
obtained by the Chandra X-ray Observatory. The image was rebinned
by a factor of 12 and smoothed to match the smoothed HRI image
in the left panel.
 \label{fig:fig2}}
\end{figure*}

\end{document}